\documentclass[sigconf,anonymous=false]{acmart}

\usepackage{algorithmic}
\usepackage{graphicx}
\usepackage{textcomp}
\setcopyright{none}
\renewcommand\footnotetextcopyrightpermission[1]{}
\usepackage{footnote}
\usepackage{xcolor}
\usepackage{booktabs}
\usepackage{makecell}
\usepackage{svg}
\usepackage{tikz}
\usepackage{multicol}
\usepackage{multirow}
\usepackage{scalerel}
\usepackage{pifont} 
\PassOptionsToPackage{hyphens}{url}\usepackage{hyperref} 
\usepackage{listings}
\AtBeginDocument{%
  \providecommand\BibTeX{{%
    \normalfont B\kern-0.5em{\scshape i\kern-0.25em b}\kern-0.8em\TeX}}}

\copyrightyear{2024}
\acmYear{2024}
\setcopyright{acmlicensed}\acmConference[FPGA '24]{Proceedings of the 2024 ACM/SIGDA International Symposium on Field Programmable Gate Arrays}{March 3--5, 2024}{Monterey, CA, USA}
\acmBooktitle{Proceedings of the 2024 ACM/SIGDA International Symposium on Field Programmable Gate Arrays (FPGA '24), March 3--5, 2024, Monterey, CA, USA}
\acmDOI{10.1145/3626202.3637578}
\acmISBN{979-8-4007-0418-5/24/03}

\acmConference[ISFPGA '24]{the 32nd ACM/SIGDA International Symposium on Field-Programmable Gate Arrays}{March 3 – March 5, 2024}{Monterey, CA}
\newcommand\copyrighttext{%
  \footnotesize \textcopyright Mark Klaisoongnoen 2024. This is the author's version of the work. It is posted here for your personal use. Not for redistribution. The definitive version was published in the 2024 ACM/SIGDA International Symposium on Field Programmable Gate Arrays, https://doi.org/10.1145/3626202.3637578}
\newcommand\copyrightnotice{%
\begin{tikzpicture}[remember picture,overlay]
\node[anchor=south,yshift=10pt] at (current page.south) {\fbox{\parbox{\dimexpr\textwidth-\fboxsep-\fboxrule\relax}{\copyrighttext}}};
\end{tikzpicture}%
}

\begin{document}


\title[Evaluating Versal AI Engines for option price discovery in market risk analysis]{Evaluating Versal AI Engines for option price discovery\\in market risk analysis}
\author{Mark Klaisoongnoen}
\orcid{0000-0001-9279-0326}
\affiliation{%
  \institution{EPCC, The University of Edinburgh}
  \city{Edinburgh}
  \country{United Kingdom}
}
\email{mark.klaisoongnoen@ed.ac.uk}

\author{Nick Brown}
\orcid{0000-0003-2925-7275}
\affiliation{%
  \institution{EPCC, The University of Edinburgh}
  \city{Edinburgh}
  \country{United Kingdom}}

\author{Tim Dykes}
\orcid{0000-0002-0401-4587}
\affiliation{%
  \institution{HPC/AI EMEA Research Lab, \protect\\Hewlett Packard Enterprise}
  \city{Bristol}
  \country{United Kingdom}
}

\author{Jessica R. Jones}
\orcid{0009-0003-7579-7245}
\affiliation{%
  \institution{HPC/AI EMEA Research Lab, \protect\\Hewlett Packard Enterprise}
  \city{Bristol}
  \country{United Kingdom}
}

\author{Utz-Uwe Haus}
\orcid{0000-0001-7292-9984}
\affiliation{%
  \institution{HPC/AI EMEA Research Lab, \protect\\Hewlett Packard Enterprise}
  \city{Zurich}
  \country{Switzerland}
}

\renewcommand{\shortauthors}{Klaisoongnoen et al.}

\begin{abstract}
Whilst Field-Programmable Gate Arrays (FPGAs) have been popular in accelerating high-frequency financial workload for many years, their application in quantitative finance, the utilisation of mathematical models to analyse financial markets and securities, is less mature. {Nevertheless, recent work has demonstrated the benefits that FPGAs can deliver to quantitative workloads, and in this paper, we study whether the Versal ACAP and its AI Engines (AIEs) can also deliver improved performance.}

{We focus specifically on the industry standard Strategic Technology Analysis Center's (STAC\texttrademark) derivatives risk analysis benchmark STAC-A2\texttrademark. Porting a purely FPGA-based accelerator {STAC-A2 inspired market risk (SIMR) benchmark} to the Versal ACAP device by combining Programmable Logic (PL) and AIEs, we explore the development approach and techniques, before comparing} performance across PL and AIEs. {Ultimately, we found that 
our AIE approach is slower than} {a highly optimised existing PL-only} {version due to limits on both the AIE and PL that we explore and describe}.
\end{abstract}




\maketitle

\pagestyle{plain}
\copyrightnotice

\section{Introduction} \label{sec:intro}

Whilst FPGAs provide low-latency advantages for high-frequency trading, motivating developers to program them using hardware description languages (HDL) such as VHDL or Verilog, these properties are not as advantageous for quantitative finance workloads, discouraging extensive efforts in that domain. However, recent advancements both in hardware and software development ecosystems have significantly enhanced the capabilities of FPGAs increasing their usability for software developers, for instance, tools such as AMD-Xilinx's Vitis \cite{vitis_exerpience_report} enable programming of them using C or C++, {and hardened components such as AI Engines (AIEs)}. {These advancements have consequently motivated communities to (re-) explore FPGAs} for their workloads \cite{brown2021porting,yang2019fully,brown2021accelerating}.


Given the suitability of dataflow architectures to this workload, in this paper we investigate the {SIMR benchmark} on AMD Xilinx's latest generation Versal Adaptive Compute Acceleration Platform (ACAP). {In order to investigate potential benefits of combining latest generation Programmable Logic (PL) and AIEs, we explore the porting of this model to the AIEs and evaluate the resulting performance properties}.

The paper's structure is organized as follows: In Section \ref{sec:bg}, we provide a survey of related activities and previous work {porting this benchmark to FPGAs}. Subsequently, in Section \ref{sec:setup}, we present the experimental setup with problem sizes of the benchmark and Versal configuration used throughout the study. Section \ref{sec:single_kernel_perf} describes efforts in moving from the PL implementation to the Versal's AI Engines (AIE) and we focus on single kernel performance, before in Section \ref{sec:pl_integration}, we describe the challenges of integrating the AIE optimisations selected in Section \ref{sec:single_kernel_perf} with the PL and appropriate coupling. Section \ref{sec:multi_kernel_perf} reports multi-kernel performance and energy usage, exploring the outcomes of different optimisation techniques and trade-offs when adopting appropriate coupling and integration with the programmable logic (PL). Finally, Section \ref{sec:conclusions} concludes and describes lessons learnt. The contribution of this work extends beyond an efficiency-driven exploration of {the SIMR benchmark} on the Versal ACAP FPGA, where we also impart insights that can be widely applied to enhance high-performance numerical modelling on state-of-the-art FPGA platforms.

\section{Background} \label{sec:bg}

Market risk analysis involves evaluating the potential impact of price fluctuations on the financial positions held by investors and traders. Falling within the broader domain of quantitative finance, which employs mathematical models and datasets to scrutinise financial markets, these workloads are computationally intensive and demand substantial computational resources. Although the prevailing approach is to execute these models on CPUs and GPUs, there have been several activities exploring the acceleration of quantitative finance on FPGAs \cite{brown2021optimisation,inggs2015high,diamantopoulos2021acceleration}.


The Strategic Technology Analysis Center (STAC) serves as the orchestrating body for the industry-leading STAC Benchmark Council\texttrademark{}. STAC members comprise some of the largest global banks, hedge funds, proprietary trading firms, exchanges, and leading technology vendors within finance. Their membership constitutes more than 400 financial institutions and over 50 technology vendors, with STAC being instrumental in presenting standardised financial benchmark suites for common financial workloads.

Relying on the exacting specifications of the STAC Benchmark Council for official audits, STAC members build and test their software and hardware systems for rigorous examination, optimisation, and validation against these real-world benchmarks. By contrast, in our research, we select components of the benchmarks to investigate the algorithmic, performance, and energy attributes and thus suitability of FPGAs for these workloads and are not bound to those strict auditing rules and thus much more flexible with implementation details. Consequently, the work and results presented in this paper are of a research and exploratory nature and must not be compared with official STAC reports and audit results. 

\subsection{Previous acceleration of the SIMR benchmark on programmable logic (PL)}


In \cite{stac-a2-h2rc}, the authors explored the porting of the {SIMR benchmark} \cite{stac-a2-2012} to the programmable logic of the AMD-Xilinx Alveo U280 and Intel Stratix-10 FPGAs, exploring the implementation differences between both architectures and the performance advantages of their host-device data streaming approach with both vendors. Building on a previous version, \cite{heart-tsukuba-stac-a2}, {Figure \ref{fig:dataflowdesign} sketches the dataflow design of the benchmark that was ported to the PL }{in \cite{heart-tsukuba-stac-a2} and which contains a detailed description of the algorithm and implementation.} {The boxes represent HLS dataflow stages and the arrows} {represent} {streams. Each stage contains three nested loops, the outer looping over assets (the number of financial options that are being analysed), the middle looping over timesteps (which determines the length of time to model the option over) and the inner loop operates over paths (which drives the accuracy). Each (inner) iteration results in the calculation of an element. 

\begin{figure}[h]
  \centering
  \includegraphics[width=0.5\columnwidth]{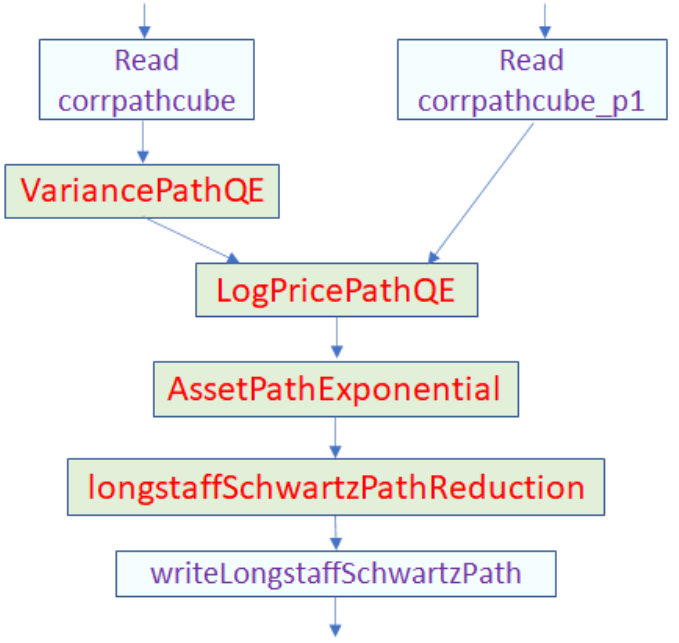}
  \caption{{Illustration of the dataflow design where stages are running concurrently connected via HLS streams \cite{stac-a2-h2rc}}}
  \label{fig:dataflowdesign}
\end{figure}

\subsection{The Versal Adaptive Compute Acceleration Platform (ACAP) and AI engine (AIE)} \label{sec:versal}

The Versal ACAP combines programmable logic (PL) with Very Long Instruction Word (VLIW) vector processors known as AI Engines (AIEs) \cite{hotchips2019_versal_ai_core} {which are capable of executing seven instructions per cycle (a scalar operation, up to two moves, two 256-bit vector loads, one 256-bit vector store and one vector manipulation operation). These are programmed} in C++ using {bespoke} calls relying on Single Instruction Multiple Data (SIMD) intrinsic data types and vector intrinsic functions. {The programming model is that of a dataflow graph, with individual kernels (that will be mapped to the AIEs) connected and data flowing between them. Code is compiled using AMD-Xilinx's \texttt{aiecompiler}, which is part of Vitis, and simulators are also provided.}

The compiler produces an \textit{Executable and Linking Format (ELF)} file to be run on the AIE processor with the resulting adaptive data flow (ADF) graph application consisting of nodes representing compute kernel functions, and edges representing data connections. For any application, the generated graph has to be controlled and marshalled for instance to initialise, run, wait or end computation. With the AIE, graphs can be specified to accommodate freely running kernels that run continuously and that will start and stop, hence restart, automatically after every last iteration has been completed removing the need to handle anything from the host side. A disadvantage of freely running AIE kernels is the fact that kernels only support streaming interfaces.

The tooling provides several options for data communication between {the kernels, which will be mapped to AIEs}. These include streams, a uni-directional cascade stream and windows. Several trade-offs exist between these communication approaches, such as the AIEs are only able to read and write 32 bits per cycle from a stream, 256 bits per cycle from windows and up to 384 bits per cycle from cascade stream. {Driven by the limits of the architecture,} there are up to two inputs and outputs per kernel, any of which can be streams or windows. Whilst the 384-bit wide cascade stream provides higher bandwidth {than a normal stream}, it is uni-directional and connects the top left corner AIE on the AIE array with the bottom right cornered AIE in a single direction, {thus imposing constraints on the mapping of kernels to AIEs.}

Windows provide memory-mapped buffers {and these work in iterations, with data being filled or consumed per iteration of the AIE graph.} Whilst filling input data windows with data before the kernel starts adds some initial latency compared to streams, {they can be used to provide a ping-pong buffer which fills and consumes concurrently, and data can be read from or written to windows in an arbitrary order, unlike streams. In this work, we concentrate on Single Precision (SP) Floating Point (FP) arithmetic, and each AIE is capable of operating on a vector width of eight SP FP numbers per cycle. Using AIEs to accelerate a computationally intensive CFD workload was investigated in \cite{brown2023exploring} where, due to the number of inputs exhausting the streams between PL and AIEs, they were limited in scaling across the AIE array, however the algorithm explored here only requires two inputs and-so this will not be a limitation.} 

\section{Experimental setup} \label{sec:setup}

For this work, we utilise an AMD-Xilinx VCK5000 featuring a Versal VC1902 ACAP with 16GB of DDR4-DRAM. All VCK5000 runs are built using Vitis 2022.1, with the PL operating at 300MHz, and the VC1902 containing 400 AI Engines running at 1.2GHz. For comparison, we use an Alveo U280 equipped with 8GB of High Bandwidth Memory (HBM2), also running at 300MHz, and Alveo kernels are built using Vitis 2021.1. Both the VCK5000 and Alveo U280 are PCIe-based cards hosted by a machine featuring a 32-core AMD EPYC 7502 processor and 256GB DRAM. 
Throughout the evaluation, all reported results are averaged over five runs. 

{Our implementation of the SIMR benchmark differs from the official STAC-A2 benchmark as follows:}
\begin{itemize}
    \item {SIMR is in single floating-point precision (STAC-A2 requires double floating-point precision),}
    \item {our implementation is without vendor involvement in optimisation,}
    \item {the code was not reviewed by STAC and}
    \item {we use problem sizes different from STAC-A2.}
\end{itemize}

Table \ref{tab:prob_sizes_defined} shows the benchmark problem sizes used throughout this work where problem sizes range from \textit{Tiny} to 
\textit{Large}. As described in Section \ref{sec:bg}, the number of paths \textit{P} defines the accuracy, and is constant across all our problem sizes. The number of assets \textit{A} represents the number of financial options that are analysed, and timesteps \textit{T} determines the option tenure ranging from half a year (126 trading days) to two years (504 trading days). 

\begin{table}[h]
  \begin{center}
  \caption{Problem Sizes with defined number of assets ($A$), time steps ($T$) and paths ($P$). The number of elements is the product of $A * T * P$. Each element requires two data points. Each data point is a 32-bit floating-point number.}
  \label{tab:prob_sizes_defined}

  \resizebox{0.8\columnwidth}{!}{ 
  \begin{tabular}{c|ccc|ccc}
    \toprule
    \makecell{\textbf{Problem} \\ \textbf{size}} & \makecell{\textbf{A}} & \makecell{\textbf{T}} & \makecell{\textbf{P}} & \makecell{Elements \\ $\left(\times 10^6\right)$} & \makecell{Datapoints \\ $\left(\times 10^6\right)$} & \makecell{Size \\ (MB)} \\
    \midrule
    Tiny         & 5  & 126 & 25 k  & 15.75 & 31.5 & 126 \\
    Small        & 10  & 126 & 25 k  & 31.5 & 63 & 252 \\
    Medium       & 20 & 252 & 25 k  & 126 & 252 & 1008 \\
    Large        & 30 & 504 & 25 k  & 378 & 756 & 3024 \\
  \bottomrule
\end{tabular}
}
  \end{center}
\end{table}

\section{Single AIE graph development} \label{sec:single_kernel_perf}

In this work, we focus on the \textit{VariancePathQE} stage of SIMR \cite{stac-a2-h2rc} exclusively as this is responsible for the largest portion of runtime and contains arithmetic operations that are all supported by the AIE tooling. In this section, we explore the porting and optimisation of the arithmetic operations executed by the \textit{VariancePathQE} stage onto the AIEs. There are a total of 36 scalar operations that must be executed by this stage to calculate each element (i.e. for each path, of each timestep, of each asset).


During development, we enabled \textit{printf} debugging (via the compiler flag \mbox{\textit{--profile}}) and captured performance numbers in cycles by accessing the cycle counter of each AIE tile as illustrated in Listing \ref{lst:print_perf_cycles}. These are the cycle numbers reported in Table \ref{tab:single_aie_kernel_cycles} and referred to throughout this section, however, we found that during hardware emulation the \textit{aiesimulator} executes individual kernels in sequential graph order, and therefore performance cycles are useful for evaluating individual kernels for stream stalls and lock stalls but require interpretation for estimating overall graph performance. 

\begin{lstlisting}[basicstyle=\fontsize{8}{9}\selectfont,frame=lines,caption={Illustration of the AIE cycle counter to capture performance data}, label={lst:print_perf_cycles}, numbers=left]
void v1qe_fn(...) {
    unsigned cycle_num[2]; 
    aie::tile tile=aie::tile::current(); cycle_num[0]=tile.cy
cles(); 
    ...
    cycle_num[1]=tile.cycles(); printf("start=%d,end=%d,total=%d\n",
    cycle_num[0],cycle_num[1], cycle_num[1]-cycle_num[0]);
}
\end{lstlisting}


{The first version ported to the AIEs comprised a single AIE kernel} with 32-bit input streams. {Containing all the operations within a single kernel, and leveraging scalar (non-vectorised) operations only for each of the required 36 calculations per element,} the performance of this is reported by the row \textit{naive scalar kernel} in Table \ref{tab:single_aie_kernel_cycles}. {This table reports the total number of cycles, the average number of cycles per kernel, and the average number of vectorised operations undertaken per cycle (with 8 being the theoretical maximum).}

\begin{table}[htb]
    \caption{Single AIE graph performance in cycles. \textit{Total cycles} is the difference between \textit{End} cycle and \textit{Start} cycle, one kernel per AIE. Procedure to capture cycles as illustrated in Listing \ref{lst:print_perf_cycles}, Utilisation \textit{U}, Number of kernels \textit{\#K}, {number of operations per cycle (\textit{Ops/cycle}) and operations efficiency (\textit{Eff}) in percentage, totalling 18,144 operations and eight operations per cycle (100\% efficiency).}}
    \label{tab:single_aie_kernel_cycles}
    \centering
    
    \resizebox{\columnwidth}{!}{ 
    \begin{tabular}{c|c|c|c|c|c|c|c} 
    \toprule
    \multirow{2}{*}{\textbf{Description}} & \textbf{U} & \multicolumn{6}{c}{\textbf{Performance}} \\

    \cmidrule{2-8}
    
    & {\#K}  &  \makecell{Start\\cycle}  &  \makecell{End\\cycle}  &  \makecell{Total\\cycles}  & \makecell{Avg. cycles/\\kernel} & \makecell{{Ops}\\{/cycle}} & \makecell{{Eff}\\ {(\%)}} \\ 
    
    \midrule
    
    \makecell{naive scalar kernel} & 1 & 615 & 642,207 & 641,592 & 641592.0 & 0.03 & 0.4 \\ 
    \makecell{vectorised kernel}  & 1 & 615 & 19,222 & 18,607 & 18607.0 & 0.98 & 12.2 \\ 
    \makecell{multi-AIE} & 9 & 1,036 & 12,465 & 11,429 & 1269.9 & 1.59 & 19.8 \\ 
    \makecell{read 128 bit/cycle for loads}  & 10 & 1,080 & 11,977 & 10,897 & 1089.7 & 1.67 & 20.8 \\ 
    \makecell{one vec operation/kernel}  & 39 & 2,453 & 30,747 & 28,294 & 725.5 & 0.64 & 8.0 \\ 
    \bottomrule
    
    \end{tabular}}
    
\end{table}

{Whilst a useful baseline, our initial naive kernel operated in scalar only and thus did not take advantage of the 8-way vectorisation for SP FP. This required leveraging the bespoke AMD Xilinx vectorisation intrinsics, such as \texttt{aie::add} and \texttt{aie::mul}, and significant restructuring of the code so that a single instruction can operate on eight floating-point elements per vector. Due to lack of support by the tooling, and availability of data, it was only possible to convert 24 of the calculations per element into vector equivalents, with 12 remaining scalar.}


In Table \ref{tab:single_aie_kernel_cycles}, the \textit{vectorised kernel} row reports performance when modifying the code to leverage 8-way vectorisation, still in a single AIE kernel. {Whilst this increases utilisation, and significantly improves performance compared to the scalar version of the code, it does add additional complexity.} For instance, additional padding on the PL side will be required for cases when the number of input elements is not dividable by eight. The input padding is required as the AIE only starts processing once the first batch of eight input elements is available to the AIE entry point kernel. {The reader might observe that the performance difference in Table \ref{tab:single_aie_kernel_cycles} between the vectorised and scalar kernel versions is greater than eight, this is because pipelining of the loops on the AIEs was also enabled by adding the \textit{chess\_prepare\_for\_pipelining} pragma directive.} 


{However, considering that there are 400 AIEs on the VC1902, leveraging only a single AIE does not make the best use of the resources. Consequently, we then further modified the code so that it could leverage multiple AIEs. The PL code that we are porting to the AIE comprises nine functions, and-so this was the basis of our AIE design}, decomposed into nine separate kernels, forming a pipeline which drives the execution of the AIEs in a graph fashion. The objective was that, after the pipeline has been filled, overall the AIE graph running across the AIEs then yields one output result vector of eight elements per cycle. {This adds an additional 20 read and write operations between the AIE kernels}, and the performance for this implementation is reported as \textit{multi-AIE} in Table \ref{tab:single_aie_kernel_cycles}, which can be observed reduces the total number of cycles further. 


{The interface between the PL and AIEs comprises 32-bit streams, but four of these can be bundled together and provide 128 bits on each access. This is useful due to the clock frequency mismatch between PL and AIEs, enabling four elements to be communicated by the slower PL per PL cycle. To this point, we were using a single input to the kernel (of 128 bits), and by splitting this into two inputs per core we doubled the bandwidth from reading 128 bits every four AIE cycles to 256 bits every four AIE cycles}. These two input streams must then be combined into a single window or cascade stream for which a separate \textit{load} kernel has been implemented. This procedure has been used for the inputs \textit{V} and \textit{Z} and is reported by the row \textit{reading 128 bits per cycle for loads} in Table \ref{tab:single_aie_kernel_cycles} (and the approach also adopted by subsequent rows). {The AIE graph of this version is illustrated in Figure \ref{fig:aie_graph} where the individual kernels are each mapped exclusively to AIEs, and connected through windows (we experiment with connecting via streams or windows in Section \ref{sec:single_AIE_graph_perf}). To start processing the graph requires inputs \textit{v} and \textit{z}, and generates the output \textit{v1qe}.} 

{Whilst in this section we are reporting results from running via simulation, rather than on the actual AIEs integrated with the PL, this technique still demonstrated some benefit which will be greater on the actual hardware when coupled in Section \ref{sec:pl_integration}}.

{To this point, individual AIE kernels leverage more than a single vector operation, which potentially stalls the pipeline graph of kernels on the AIEs, and-so we explored placing a single operation in each kernel, resulting in 39 AIE kernels for the whole AIE graph. This is described by the row \textit{one vector operation per kernel} in Table \ref{tab:single_aie_kernel_cycles}.} As we maintain a \textit{runtime ratio}\footnote{{The runtime ratio defines how much of the processing time of a single AI Engine core is required by the respective kernel.}} of 1 for all kernels, every kernel is still assigned and bound to a separate AIE. For this version, the code is decomposed from its original structure in AIE kernels that contain individual arithmetic operations. It should be highlighted that the mapping of input and output ports for the graph, between kernels and for the entry point and final stage yields significant boilerplate code for the graph specification. As an example, the first two versions \textit{naive scalar kernel} and \textit{vectorised kernel} in Table \ref{tab:single_aie_kernel_cycles} resulted in tens of lines of code however the final \textit{one vector operation per kernel} variant in Table \ref{tab:single_aie_kernel_cycles} required hundreds of lines of code, adding significant complexity.

One challenge was that, as of Vitis 2022.1, the tooling supports FP division only when operating with scalars when using \textit{aie::div}. Consequently, we implemented our own vector division using the inverse \textit{aie:inv} followed by multiply \textit{aie::mul} operations, both of which are supported for SP FP vectors. At the time of writing, there is also no support for the exponential arithmetic operation on vector types. Therefore, we implement previous stages of the \textit{variancePathQE} function that builds on an intermediate result of \textit{expf{}}, which we offload to AIE, on the programmable logic of the Versal and feed the results as input to the corresponding AIE kernel.


\begin{figure*}[]
    \centering
    \vstretch{1.7}{\includegraphics[width=\textwidth]{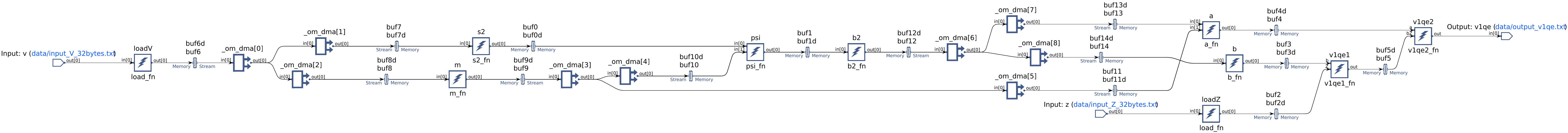}}
    \caption{Graph for \textit{multi-AIE} with ten kernels as presented in Table \ref{tab:single_aie_kernel_cycles}.}
    \label{fig:aie_graph}
\end{figure*}


\section{Coupling AI Engines with the PL} \label{sec:pl_integration}
\label{sec:single-pl}
{Section \ref{sec:single_kernel_perf} explored porting the \textit{VariancePathQE} kernel of the {SMIR benchmark \cite{stac-a2-h2rc}} onto the AIEs, but this must be integrated with the existing parts of the code running on the PL which, as it can be seen from Figure \ref{fig:dataflowdesign}, require the sending of input data to \textit{VariancePathQE} and integrating returned results from the AIEs with the subsequent dataflow stages \textit{LogPricePathQE, AssetPathExponential, longstaffSchwartzPathReduction and writeLongstaffSchwartzPath}. A challenge is that developing for the AI Engine architecture fundamentally differs from that of the reconfigurable PL and these must be integrated in a way that the }{mismatched} {clock frequencies between the PL and AIE array can be overcome. For instance by leveraging higher bandwidth per cycle on the PL as described in Section \ref{sec:single_kernel_perf}. Furthermore, whilst hardware build times for the PL can be very long, these are far shorter (seconds or minutes) for the AIEs, with a \textit{re-packaging} option to avoid rebuilding the entire bitstream if the AIE code alone has changed.} {The runs reported in Section \ref{sec:single_kernel_perf} for our AIE design were using the AIE simulator, and when coupling with the PL code described in Section \ref{sec:pl_code}, some tweaks are required.}

\subsection{{Programmable logic (PL) code}} \label{sec:pl_code}

{A challenge with the FPGA code 
is that the PL uses buffers in a nested loop as shown in Listing \ref{lst:cached_buf}. This results in a circular dependency, which can severely limit performance on the FPGA but, due to \emph{path} being the inner loop and there being a large number of paths, we can ensure} that by the time a specific value is required from the \emph{cached\_buf} array, it has been calculated and written. 
Previously our approach used an \emph{HLS dependence} pragma statement on line 5 to signify a false dependency for \texttt{cached\_buf} between iterations. 

\begin{lstlisting}[basicstyle=\fontsize{8}{9}\selectfont,frame=lines,caption={Cached buffer in nested loops}, label={lst:cached_buf}, numbers=left]
dtype cached_buf[SIZE], result;
asset_loop: for (unsigned int asset = 0; asset < assets; asset++) {
  timestep_loop: for (unsigned int i=0; i<timesteps+1; i++) {
    path_loop: for (unsigned int path=0; path<path_size; path++) {
#pragma HLS dependence variable=cached_buf inter false
      ... 
      result = arithmetic_ops(..., cached_buf[path]);
      cached_buf[path] = result;  ...
}}}
\end{lstlisting}





However, this algorithmic pattern, where an iteration requires data from a previous iteration, is a problem on the AIEs because loops are not allowed in the AIE dataflow graph. Put simply, the AIE graph model does not support connecting the output of a graph iteration as input to the next graph iteration, and-so we must transfer data between the AIE array and PL in a \textit{loopback} fashion. Consequently, to handle this we introduced a \textit{aie\_loopback\_adaptor} HLS kernel {which provides support on the PL to cache result data from the AIEs and feed this back as an input to subsequent iterations as required. This is illustrated in Figure \ref{fig:loopback}, where there are two dataflow regions connected with an internal stream. The \textit{Handle return data} stage reads results from the AIE array and sends them to the other stage and also to the PL kernel. The second, \emph{Loopback function} stage receives results from the previous stage (as long as it is not the first timestep, as otherwise initial values are provided by the PL kernel), stores these in on-chip BRAM or URAM, and serves the cached version required for the current path, timestep and asset. Two separate dataflow regions are required due to the interaction between the AIE array and PL kernel, as otherwise, this could stall a single region.} 

\begin{figure}[h]
  \centering
  \includegraphics[width=0.7\columnwidth]{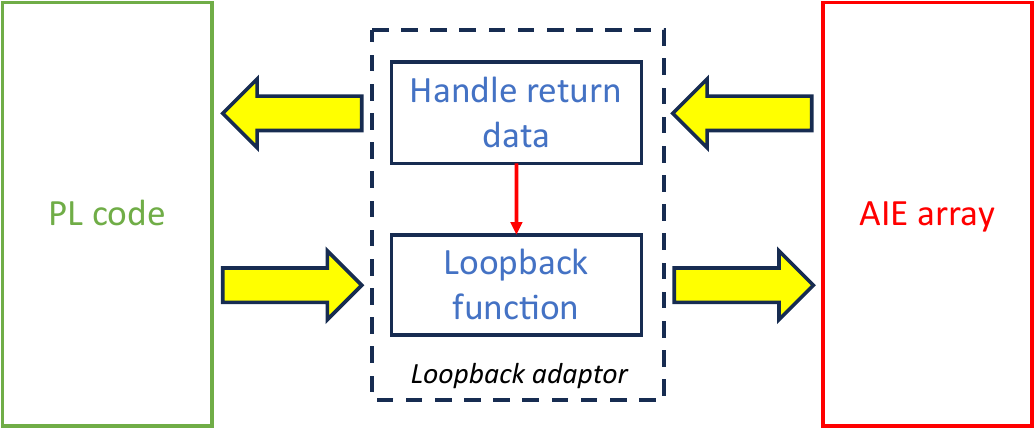}
  \caption{{Illustration of the loopback adaptor on the PL used to support the cycling of data between AIE iterations}}
  \label{fig:loopback}
\end{figure}

\subsection{{Single AIE graph performance}}
\label{sec:single_AIE_graph_perf}

Table \ref{tab:single_kernel_perf} reports the performance results\footnotemark  of our single-PL-kernel experiments on the VCK5000. Moving {the SIMR benchmark} to the VCK5000 we report the performance for the version built in pure PL without any AIEs as \textit{VCK5000 PL-only}. As the VCK5000 only provides relatively slower DRAM compared to the U280, it can be seen that performance on the VCK5000 is penalised where for the \textit{tiny} problem size the runtime has more than doubled. This performance gap between both PL-only versions on the different FPGAs decreases when scaling the problem size up to \textit{large} with the U280 still outperforming the VCK5000 by more than 1.25$\times$.

\footnotetext{The experiments conducted have not been designed to comply with STAC benchmark specifications. Therefore the experimental results that we present are of a research nature and are not representative of official STAC benchmark results.}

Table \ref{tab:single_kernel_perf} reports the performance of our kernels implemented across both PL and AIEs using the same naming and description as in Table \ref{tab:single_aie_kernel_cycles}. It can be seen that the \textit{naive scalar kernel} version, implemented with streams reading and writing in 32 bits per cycle, yields the longest total execution time as this does not use the AIEs' vectorisation capabilities, on a single AIE resulting in a more than 208$\times$ overhead compared to the \textit{VCK5000 PL-only} implementation. 

For the \textit{vectorised kernel} version in Table \ref{tab:single_kernel_perf}, we report the performance implemented in 32-bit streams and alternatively with input and output windows of 128 bytes. The windows-based version with a window size of 128 bytes performs marginally better, but both versions are slower than running on the PL alone. One reason is that even with 128 bytes of window size the memory-mapped data communication mechanism does not significantly perform better than the 32-bit stream variant.

Increasing the utilisation on the AIE array, the \textit{multi-AIE} version is the \textit{reading 128 bit per cycle for loads} varient from Table \ref{tab:single_aie_kernel_cycles} which splits the single vectorised kernel into multiple smaller kernels, each comprising multiple vector operations. Table \ref{tab:single_kernel_perf} reports performance when connecting with streams or 128 byte windows, and it can be seen that using windows delivers best performance for the \textit{multi-AIE} version across all problem sizes. The \textit{multi-AIE} version with windows is around 2$\times$ faster than the \textit{vectorised kernel} version with streams for the \textit{large} problem size but still substantially slower than the PL-only version.

The \textit{one vector operation per kernel} row, with 39 AIEs, provides performance that is marginally worse for the \textit{tiny} problem size compared to the fastest \textit{multi-AIE} version. However, when scaling the problem size this performance gap increases, with \textit{multi-AIE} performing up to 1.57$\times$ better on the \textit{large} problem size. Whilst we are able to increase the utilisation with a single graph instance, when profiling it was found that the many connections between these 39 AIEs increased the stall frequency. With overlapping dependencies in the graph as depicted in Figure \ref{fig:aie_graph}, tweaking one kernel to reduce lock or memory stalls resulted in stalls in other kernels making it difficult to balance latency across the whole graph thus not fully leveraging the available compute across AIEs.

{Ultimately, irrespective of the version, the AIEs fail to match the performance of running over the PL alone and there are several reasons for this. Firstly, the \textit{multi-AIE} version, which is best performing, contains multiple vectorised operations per kernel and-so does not provide a perfect pipeline generating results each cycle. When moving to one operation per kernel, the communication overhead between AIEs becomes dominant causing a significant number of memory stalls. Furthermore, the loopback adaptor on the PL, adds additional complexity and is a source of stalling.}

\begin{table*}[]
    \centering
    \caption{\textit{Single Kernel} and \textit{Multi Kernel} performance across PL and AIE with total execution time in ms, averaged over five runs, \textit{Connection mode} with either \textit{stream} or 128-byte AIE \textit{window}, \textit{stream} for AIE designs fully implemented with streams, \textit{vs PL} in times of VCK5000 PL-only runtime.} 
    \label{tab:single_kernel_perf}
    
    \resizebox{0.7\textwidth}{!}{ 
    \begin{tabular}{c|c|c|cc|cc|cc|cc}
        \toprule
        & \textbf{Description} & \makecell{\textbf{Connection mode}} & \textbf{Tiny} & vs PL & \textbf{Small} & vs PL & \textbf{Medium} & vs PL & \textbf{Large} & vs PL \\ 
        
        \midrule

        \parbox[t]{2mm}{\multirow{8}{*}{\rotatebox[origin=c]{90}{Single Kernel}}}  &  Alveo U280 
        &   -   & 79.29 & 0.48 & 132.44 & 0.68 & 474.31 & 0.79 & 1370.13 & 0.80 \\ 
        &  VCK5000 PL-only   &   -   & 164.75 &  -  & 194.02 &  -  & 600.55 &  -  & 1718.93 &  - \\
        
        \cmidrule{2-11}
        
        & naive scalar kernel  &  stream  & 34366.74 & 208.60 & 68711.55 & 354.15 & 274903.37 & 457.75 & 819422.49 & 476.70 \\ 
        & vectorised kernel  &  stream  & 367.78 & 2.23 & 698.68 & 3.60 & 2709.01 & 4.51 & 8040.33 & 4.68 \\ 
        & vectorised kernel  & window & 361.22 & 2.19 & 690.15 & 3.56 & 2678.54 & 4.46 & 7961.03 & 4.63 \\ 
        & multi-AIE  &  stream  & 328.18 & 1.99 & 587.32 & 3.03 & 2309.43 & 3.85 & 6573.25 & 3.82 \\ 
        & multi-AIE  & window & 285.63 & 1.73 & 511.03 & 2.63 & 1531.56 & 2.55 & 4094.08 & 2.38 \\  
        & one vec op per kernel  & window & 298.83 & 1.81 & 555.72 & 2.86 & 2153.28 & 3.59 & 6437.52 & 3.75 \\ 

        \midrule
        \midrule



        \parbox[t]{2mm}{\multirow{5}{*}{\rotatebox[origin=c]{90}{Multi Kernel}}}  
        &  Alveo U280 - 6 CU 
        &   -   & 77.81 & 0.46 & 98.46 & 0.37 & 221.71 & 0.54 & 544.48 & 0.58 \\
        
        & VCK5000 PL-only - 10 CU  &  -  & 170.11 & - & 266.38 & - & 412.16 & - & 932.76 & - \\ 
       
        \cmidrule{2-11}
         
        & 6 PL-CU + 6 AIE subgraphs &  stream  & 336.88 & 1.98 & 521.17 & 1.96 & 1547.06 & 3.75 & 3198.02 & 3.43 \\ 
        & 6 PL-CU + 6 AIE subgraphs & window & 307.38 & 1.81 & 503.09 & 1.89 & 1543.63 & 3.75 & 3193.02 & 3.42 \\ 

        \bottomrule
    \end{tabular}
    }

\end{table*}


\section{Multi AIE graph performance} \label{sec:multi_kernel_perf}
In the results reported in Section \ref{sec:single-pl} the AIEs failed to match the performance of the PL on either the VCK5000 or Alveo U280. However, we were only using a small number of AIEs and hence the question was how might this performance be impacted when the number is increased. We therefore undertook experiments with multiple kernels, and Table \ref{tab:single_kernel_perf} reports this multi-kernel performance on the PL and integrated with the AIEs across the defined problem sizes ranging from \textit{tiny} to \textit{large}. 


In \cite{stac-a2-h2rc}, we were able to fit six CUs on the Alveo U280. {By contrast, on the VCK5000 the 16.3 MB of URAM on-chip memory is large enough for our \textit{large} problem size to fit 16 CUs, but the 16 GB of DRAM for the overall problem input size limits the actual number of CUs to ten and this is reported shown in Table \ref{tab:single_kernel_perf}.} Compared to the Alveo U280 with 6 CUs, the {\textit{VCK5000} version results in total execution between 2.1$\times$ and 1.7$\times$ slower for the \textit{tiny} to \textit{large} problem sizes respectively.} Building on the host-device data transfer approach from \cite{stac-a2-h2rc} that utilises the larger off-chip memory as a streaming substrate, the VCK5000 is penalised by slower DRAM even with ten CUs compared to the same streaming approach on the Alveo U280 and its HBM with six CUs.

The \textit{multi-AIE} version in Table \ref{tab:single_kernel_perf} leverages six CUs on the PL, each CU being separately served with its own subgraph on the AIE. {The Vitis tooling reports that the available control master interfaces on the PL are exhausted beyond six CUs and consequently, with 10 kernels on the AIEs per subgraph, and each of these bound to an exclusive, single AIE, this results in a total utilisation of 60 AIEs which is 15\% of the AIE array. Therefore whilst one might attempt to overcome the performance limits of a single AIE graph discussed in Section \ref{sec:single_AIE_graph_perf} by scaling out, this is only possible to a limited extent. This issue is because the number of AXI ports required on the PL becoming too large, 14 are required per CU and with six CUs the tooling can accommodate 84 but no more.} Due to this limitation we explored coupling multiple AIE graphs to a single CU, as this can then drive a greater portion of the AIE array. However, after calculation on the AIEs, the results must be reduced on the PL with dependencies between paths. Due to this dependency, the reduction is still operating in a scalar fashion and that limits data streaming performance to and from the multiple AIE graphs.

The \textit{PL-only} approach is between 1.81$\times$ and 3.75$\times$ faster than the window based \textit{multi-AIE} version, and 1.96$\times$ to 3.75$\times$ faster than the stream based \textit{multi-AIE} version. 

\section{Conclusions} \label{sec:conclusions}

In this paper, we focused on porting the {SIMR} benchmark to the Versal ACAP combining programmable logic, and specifically the \emph{VariancePathQE} kernel to the AIEs. Focusing on the AIE development process, we investigated the data connections between AIEs, techniques for minimising stalling on individual AIE kernels, vectorisation of operations and the requirements to make data available fast enough to the compute. It was observed that moving to a higher utilisation on the AIE array is beneficial in reaching a lower average number of cycles per kernel, but this increases the complexity when balancing latency across kernels for graphs with competing or dependent branches.

{Integration with the PL was a challenge due to the cyclical nature of the algorithm, where results from one iteration are required subsequently as input. Because cycles are disallowed by the AIE graph we developed a \emph{loopback adaptor} on the PL to cache and serve intermediate results.} Consequently, the uni-directional flow of the AIE graph is well-suited for applications that can be mapped onto such a graph but application requirements such as circular connections of the graph requires further engineering efforts.

When scaling from the single AIE graph to {multiple graphs on the AIE array}, we found that 
ultimately it was not possible to match the performance of the PL-only approach {although this is an optimised dataflow design specifically tuned for the PL}. {There were four main reasons which limit performance; 1) our best performing AIE design has multiple vector operations per kernel, thus the design is unable to operate as a perfectly efficient pipeline and which generates a result each cycle, 2) when refactoring to one vector operation per kernel the design is dominated by inter-AIE communication overheads and memory stalls, 3) the loopback adaptor on the PL adds additional complexity and overhead that can stall the AIEs, 4) due to limits in the number of AXI interfaces on the PL we are unable to scale beyond 15\% of the AIEs on the array, thus meaning it is not possible to ameliorate single kernel performance by scaling out.}


\newpage
\begin{acks}
The authors express their gratitude to HPE who funded this work via their EMEA Research Lab internship programme and to STAC for granting access to the STAC-A2 benchmark and providing valuable advice and support. Additionally, we extend our acknowledgement to the ExCALIBUR H\&ES FPGA and CGRA testbeds and the AMD-Xilinx HACC program for generously providing the compute resources utilised in this research. 
\end{acks}

\newpage
\bibliographystyle{ACM-Reference-Format}
\bibliography{bib/stac}

\end{document}